\documentclass{ws-ijgmmp}

\usepackage{bm}
\usepackage{stackrel}
\usepackage{comment}
\usepackage{empheq}

\begin{document}

\markboth{T. Wada, A.M. Antonio, H. Matsuzoe}
{HUYGENS' EQUATIONS AND THE GRADIENT-FLOW EQUATIONS IN INFORMATION GEOMETRY }

%
\catchline{}{}{}{}{}
%

\title{HUYGENS' EQUATIONS AND THE GRADIENT-FLOW EQUATIONS IN INFORMATION GEOMETRY 
}

\author{TATSUAKI WADA
}

\address{ Region of Electrical and Electronic Systems Engineering, Ibaraki University, \\
4-12-1 Nakanarusawa-cho, Hitachi, Ibaraki, 316-8511, Japan\\
\email{tatsuaki.wada.to@vc.ibaraki.ac.jp
} }

\author{ANTONIO M. SCARFONE}

\address{Istituto dei Sistemi Complessi, Consiglio Nazionale delle Ricerche (ISC-CNR) c/o Politecnico di Torino, \\
Corso Duca degli Abruzzi 24, Torino, 10129, Italy \\
antonio.scarfone@polito.it}

\author{HIROSHI MATSUZOE}

\address{Department of Computer Science, Nagoya Institute of Technology, \\
Gokiso-cho, Showa-ku, Nagoya, Aichi, 466-8555, Japan \\
matsuzoe@titec.ac.jp }

\maketitle

\begin{history}
\received{(Day Month Year)}
\revised{(Day Month Year)}
\end{history}

\begin{abstract}
We revisit the relation between the gradient-flow equations and Hamilton's equations in information geometry. By regarding the gradient-flow equations as Huygens' equations in geometric optics, we have related the gradient flows to the geodesic flows induced by the geodesic Hamiltonian in an appropriate Riemannian geometry.
The original evolution parameter $\textit{t}$ in the gradient-flow equations is related to the arc-length parameter in the associated Riemannian manifold by Jacobi-Maupertuis transformation.
As a by-product,  it is found the relation between the gradient-flow equation and replicator equations.
%
\end{abstract}


\section{Introduction}\label{sec1}
More than two decades ago,
the gradient flows in information geometry (IG) were studied in the pioneering works by
Fujiwara \cite{F93}, Fujiwara and Amari \cite{FA95}, Nakamura \cite{N94}.
On a dually flat space $( \mathcal{M}_S, g,  \nabla, \nabla^{\star} )$, they define a potential function 
$U( p) := D_{\nabla}( p, q)$ at a fixed point $q$ in $\mathcal{M}_S$, where $D_{\nabla}( p, q)$ is the divergence function associated with the affine connection $\nabla$.
It is remarkable, due to the dually-flat structure in IG,  that the nonlinear gradient flow equations in the affine coordinates $\{ \theta^i \}$
are equivalent to the linear differential equations in the dual affine coordinates $\{ \eta_i \}$.
They showed that these linear differential equations 
are expressed as Hamilton's equations when the space has even dimensions. Specifically, the linear differential equations 
\begin{align}
  \frac{d \eta^{\rm gf}_j }{ dt}  = -\eta^{\rm gf}_j,  \quad j=1,2, \dots, 2 m,
  \label{orig-gradEq}
 \end{align}
 coincide with 
Hamiltonian's equations for their proposed Hamiltonian 
\begin{align}
H^{\rm gf} = - \sum_{k=1}^m Q^k  P_k, 
\end{align}
where  $Q^k =  \eta^{\rm gf}_{2k}$ denotes the generalized position and $P_k = -(1/ \eta^{\rm gf}_{2k-1})$ the generalize momentum, respectively.
Here, each $k$-th term $- Q^k P_k$, ($k=1, \ldots, m$) in the Hamiltonian $H^{\rm gf} $ is a first integral, i.e., a conserved quantity with respect to time $t$ evolution. Since the system described by Hamilton's equations for this $H^{\rm gf} $ is in an  $m$-dimensional subspace of  the $2m$-dimensional symplectic space and the system has $m$-conserved quantities, this is a completely integrable system. 
This relation between Hamilton flows $(Q^k(t), P_k(t) )$ and gradient flows $\eta^{\rm gf}_i(t)$ is mysterious and incomprehensible, because in general a Hamilton system is conservative  whereas a gradient system is often used to model a dissipative (non-conservative) system. Later,
Boumuki and Noda \cite{BN16} studied this relationship from the view point of symplectic geometry.
However the meanings of this Hamiltonian $H^{\rm gf} $ and the parameter $t$ had been unclear. Then
in our previous paper \cite{WSM21}, we studied these issues by applying the generalized eikonal equation, which is usually used in geometric optics, to a simple thermodynamic system and showed that the time parameter $t$ of the gradient flow equations in IG is related to the temperature of the simple thermodynamic systems based on the Hamilton-Jacobi dynamics.
The key point is that the time parameter $t$ of a system is conjugate  to the energy $E$ which is the value of the Hamiltonian just as the positions $\{ q^i \}$ are conjugate to the momenta $\{ p_i \}$.  
Through our previous study \cite{WSM21} we realize the importance of treating space and time on equal footing, which is an essence of Einstein's relativity \cite{Dirac}. It is known, in general relativity, that  the selection of coordinates system in curved space-time is important in order to conveniently describe physical equations. This is non-trivial and in a suitable coordinate system, the physical equations have simple forms and clear physical meanings \cite{G18}.  
In addition,  treating space and time on equal footing is important also in optics.
Indeed, Cartan pointed out in his famous book \cite{Cartan} in chap.~XIX on page 216 that ``\textit{..., even in optics, the time coordinate does not play a role that is essentially different from the one that is played by the spatial coordinates. The fundamental laws of optics are not necessarily related to the classical notions of space and time, and they behave just as they do in the theory of relativity}''.  Here ``\textit{the classical notions of space and time}'' means the concept of Newton's absolute time, in which time and space are not treated on equal footing.
Recall that historically Hamilton invented his dynamical formalism (Hamilton's equations) for a classical point particle from his theory of systems of rays \cite{Hamilton}.

Hereafter, if nothing is stated, we use Einstein's summation convention for repeated indices.
We denote vectors with $m$-components by boldface letters (e.g., $\bm{q}, \bm{p}$).
In this work, we consider the two different sets of the gradient-flow equations
\begin{align}
  \frac{d \eta_i}{dt} = - g_{ij}(\bm{\eta}) \, \frac{\partial }{\partial \eta_j} \Psi^{\star}(\bm{\eta}),
  \quad \textrm{and} \quad 
  \frac{d \theta^i}{dt} = g^{ij}(\bm{\theta}) \, \frac{\partial }{\partial \theta^j} \Psi(\bm{\theta}),
\label{gradEq_eta}
\end{align}
for the potential functions $\Psi(\bm{\theta})$ and $\Psi^{\star}(\bm{\eta})$ of a dually-flat statistical manifold $(\mathcal{M}, g, \nabla, \nabla^{\star})$,
where the relations $\theta^i = \partial \Psi^{\star}(\bm{\eta}) /  \partial \eta_j,  \eta_i = \partial \Psi(\bm{\theta}) /  \partial \theta^j, g_{ij}(\bm{\eta}) = \partial \eta_i / \partial \theta^j$ and $g^{ij}(\bm{\theta}) = \partial \theta^i / \partial \eta_i$ hold. By using these relations,
it is well known for the dually-flat structure \cite{FA95} in IG that the sets of the nonlinear differential equations in \eqref{gradEq_eta} are transformed to the following two set of the linear differential equations
\begin{align}
  \frac{d \theta^i}{dt} = - \theta^i, \quad \textrm{and} \quad \frac{d \eta_i}{dt} = \eta_i, \quad i=1,2, \ldots, m,
 \label{gradEq}
\end{align}
respectively.  For example, the first set of the gradient-flow equations in \eqref{gradEq_eta} are transformed as
\begin{align}
   \frac{d \theta^i}{dt} =  \frac{\partial \theta^i}{\partial \eta_j}  \frac{d \eta_j}{dt} = g^{ij} (\bm{\theta}) \left(  - g_{jk}(\bm{\eta}) \, \frac{\partial }{\partial \eta_k} \Psi^{\star}(\bm{\eta}) \right) = - \delta^i_k \, \theta^k = - \theta^i,
\end{align}
where $g^{ij} (\bm{\theta}) \, g_{jk}(\bm{\eta}) = \delta^i_k $ is used and $\delta^i_k$ denotes Kronecker's delta. The second set of the gradient-flow equations  in \eqref{gradEq_eta} is transformed to the second set of the  linear differential equations in \eqref{gradEq} in a similar way.
These linearizations  are due to the powerfulness of dually-flat structures in IG.

It is noteworthy that the first and second sets of equations in \eqref{gradEq_eta} describe different processes in general. Of course, the first (or second) set in \eqref{gradEq_eta} and the first (or second) set in \eqref{gradEq} describes a same process in different ways, since the $\theta$- and $\eta$-coordinates are mutually related.
If the two sets of differential equations in \eqref{gradEq} are combined and imposing the conditions $\eta^{\rm gf}_{2i} =  \theta^i$ and $\eta^{\rm gf}_{2i-1} = - 1/ \eta_i$, then the differential equations \eqref{orig-gradEq} are reproduced. These conditions lead to the $m$ conserved quantities $\theta^i \, \eta_i = -1, \textrm{(no summation)}, i=1,2,\dots, m$ in the Hamiltonian $H^{\rm gf}$.
In contrast, we consider the two sets of the gradient-flow equations as describing different processes in general.
In this work we mainly consider the  first set of equations in \eqref{gradEq} (or  in \eqref{gradEq_eta}). 
The key idea of our work is regarding the set of the gradient-flow equations in IG as Huygens' equations in geometric optics.
This leads to a natural Hamiltonian \eqref{H} with a potential function $-n(\bm{q})$, but in order to obtain  a geodesic flow we need a geodesic Hamiltonian, i.e, a Hamiltonian without any potential function. To this end one resorts to the Jacobi-Maupertuis (JM) transformation, which transform a natural Hamiltonian to a geodesic Hamiltonian.
These processes enable us to relate the gradient flow, say $\theta^i(t)$, to the geodesic flow $(q^i(\tau), d q^i(\tau)/d \tau )$  or  the co-geodesic flow $(q^i( \tau), p_i( \tau))$ induced by a geodesic Hamiltonian on an appropriate Riemannian manifold. Here $\tau$ is the arc-length parameter of this Riemannian manifold. In general, there are many different flows along the same path. For example, consider a constant velocity (say $v_0$) motion (or flow) $x(t)$ along a straight path. If one changes the time parameter $t$ to $s(t) = a \, t + b$, where $a$ and $b$ are constants\footnote{this kind of parameters is called an \textit{affine} parameter and physically means in general that a flow $x(s)$ with an affine parameter $s$ describes a constant velocity motion along its path.}, the flow $x(s)$ has the different velocity $d x(s) / ds = v_0 / a$. In this way it is important to specify a time parameter which describes a flow along the same path.
In this work, we shall finally relate the time parameter $t$ in the gradient-flow equations to the above arc-length parameter $\tau$ of an appropriate Riemannian manifold.

The rest of the paper is organized as follows.
In Section \ref{sec:2} we first give a quick review of the geometric optics for isotropic media. We explain \textit{Fermat's principle, eikonal equations, Huygens' equations}, which are complementary to each other 
in order to describe a ray path in an optical medium.  
Next in subsection 2.1, we briefly review the geometric optics for anisotropic media.
From anisotropic Huygens equation, or equivalently, the corresponding Fermat principle, we obtain the Euler-Lagrange equations for a ray path in an anisotropic medium. The corresponding Hamiltonian is also obtained and by applying  the Jacobi-Maupertuis (JM) transformation, we obtain the geodesic Hamiltonian, which describes a geodesic flow on a Riemannian manifold. 
In Section \ref{sec:3}, we propose a similar geodesic Hamiltonian in IG and study some relations with the gradient flow equations \eqref{gradEq_eta}.
As concrete examples we consider a Gaussian probability distribution function (pdf) in sub-section \ref{Gaussian} and a Gamma pdf in sub-section \ref{Gamma}. Some relations between the gradient flow equations \eqref{gradEq_eta} and the replicator equations are pointed out in \ref{RepEq}. The second set of the gradient-flow equations in \eqref{gradEq_eta} is discussed in sub-section \ref{2ndset}.
The final Section \ref{sec:conclusions} is devoted to our conclusions.
In Appendix A, Maupertuis' principle and Jacobi transformation are explained.

\section{Geometric optics}
\label{sec:2}

Here we briefly review the basic of geometric optics. 
Firstly, we consider geometric optics for isotropic media and then in the next subsection for anisotropic media.

Fermat's principle (or the principle of least time) is well known in geometrical optics \cite{Holm} and it states that the stationarity of the so-called optical length, i.e,
\begin{align}
  0 = \delta \int_A^B n(\bm{r}(s)) \, \sqrt{ \frac{d \bm{r}}{ds} \cdot  \frac{d \bm{r}}{ds} } \, ds = \delta \int_A^B n(\bm{r}(s)) \, ds ,
\end{align}
for the path $\bm{r}(s)$ of a light ray passing from a point $A$ to another point $B$ 
in a three-dimensional ($m=3$) medium with a refractive index $n(\bm{r}) := c/ v(\bm{r})$, where $c$ denotes the speed of light in vacuum and $ v(\bm{r})$ is the phase speed of light at a position $\bm{r}$ in an optical medium.
 Here $s$ denotes the arc-length parameter of the ray path $\bm{r}(s)$ and
its line element $ds$ satisfies the relation
\begin{align}
  (ds)^2 = d \bm{r}(s) \cdot d \bm{r}(s).
\end{align}
From Fermat's principle, the equation for the ray path $\bm{r}(s)$ is obtained as
\begin{align}
  \frac{d}{ds} \left( n(\bm{r}) \frac{ d \bm{r}}{ds} \right) = \frac{\partial n(\bm{r})}{\partial \bm{r}} = \nabla n(\bm{r}),
  \label{eikonal}
\end{align}
which is called the \textit{eikonal equation}.

Huygens' wavelets is complementary to Fermat's ray optics. According to Huygens' assumption for wavelets, a level set  of the wave front $S(\bm{r})$
moves along the ray vector $\bm{n}(\bm{r})$ so that its infinitesimal change along the ray over an infinitesimal distance $d \bm{r}$ is given by
\begin{align}
   \nabla S(\bm{r}) \cdot d \bm{r} = \bm{n}(\bm{r}) \cdot d \bm{r} = n(\bm{r}) \, ds.
\end{align}
This leads to so called \textit{Huygens' equation} for isotropic media
\begin{align}
   \nabla S(\bm{r}) = n(\bm{r}) \, \frac{d \bm{r}}{ds},
   \label{HuygensEq}
\end{align}
from which we see that
\begin{align}
  \left\vert \nabla S(\bm{r}) \right\vert^2 = n^2(\bm{r}) \, \frac{d \bm{r}}{ds} \cdot \frac{d \bm{r}}{ds} = n^2(\bm{r}).
  \label{nsquare}
\end{align}
Note that Fermat's eikonal equation \eqref{eikonal} is obtained from Huygens' equation \eqref{HuygensEq} as follows.
By using the chain rule and \eqref{HuygensEq}, we have
\begin{align}
  \frac{d}{ds} = \frac{d \bm{r}}{ds} \cdot \nabla = \frac{1}{n(\bm{r})} \nabla S(\bm{r}) \cdot \nabla.
\end{align}
Applying this relation to Huygens' equation \eqref{HuygensEq} leads to
\begin{align}
  \frac{d}{ds} \left( n(\bm{r}) \frac{d \bm{r}}{ds} \right) &= \frac{d}{ds}   \nabla S(\bm{r})   = \frac{1}{n(\bm{r})} \nabla S(\bm{r}) \cdot \nabla \Big( \nabla S(\bm{r})  \Big)
 \nonumber \\
 &= \frac{1}{ 2 n(\bm{r})} \nabla \left\vert \nabla S(\bm{r}) \right\vert^2   = \frac{1}{ 2 n(\bm{r})} \nabla n^2(\bm{r}) = \nabla n(\bm{r}),
\end{align}
where \eqref{nsquare} is used.
It is worthwhile to note that Fermat's principle of stationary ray paths and Huygens' principle of constructive interference of wavelets are dual to 
each other.

\subsection{Geometric optics in anisotropic media}
We next review the  geometric optics in anisotropic media \cite{Holm} with a general dimension $m$. 
Due to the anisotropy of an optical medium, the tangent vector $d \bm{q}(s) / ds $ of a ray $\bm{q}(s)$ is not normal to its wave front $\nabla S(\bm{r})$ but instead obeys
the following relation 
\begin{align}
  \frac{d \bm{q}(s)}{ds} =  \mathcal{D}^{-1}(\bm{q}) \, \nabla S(\bm{q}),
  \quad \textrm{or} \quad
   \frac{d q^i}{ds} = \mathcal{D}^{ij}(\bm{q}) \, \frac{\partial S(\bm{q}) }{\partial q^j}, \quad i, j = 1,2, \ldots, m,
\end{align}
which depends on the location along the ray path $\bm{q}(s)$.
Here $\mathcal{D}$ is an invertible matrix function characterizing the anisotropic medium of interest.
The \textit{anisotropic Huygens equation} is written by
\begin{align}
   \frac{\partial S(\bm{q}) }{\partial q^i} =  \mathcal{D}_{ij}(\bm{q}) \, \frac{d q^j}{ds}, \quad i, j = 1,2, \ldots, m,
   \label{anisoHuyEq}
\end{align}
where the matrix function $\mathcal{D}$ can be expressed as
\begin{align}
  \mathcal{D}_{ij}(\bm{q}) = n(\bm{q}) \, g_{ij}(\bm{q}).
  \label{D}
\end{align}
From Eqs. \eqref{anisoHuyEq} and \eqref{D}, it follows that
\begin{align}
  \left\vert \nabla S(\bm{q}) \right\vert^2 &=  g^{ij}(\bm{q}) \, \frac{\partial S(\bm{q}) }{\partial q^i} \, \frac{\partial S(\bm{q}) }{\partial q^j} 
  = g^{ij}(\bm{q}) \, \mathcal{D}_{ik}(\bm{q}) \frac{d q^k}{ds} \,  \mathcal{D}_{j\ell}(\bm{q}) \, \frac{d q^{\ell}}{ds} \nonumber \\
  &=  n^2(\bm{q}) g^{ij}(\bm{q}) \, g_{ik}(\bm{q}) \,  g_{j\ell}(\bm{q}) \, \frac{d q^k}{ds}  \frac{d q^{\ell}}{ds}  = n^2(\bm{q}) g_{ij}(\bm{q}) \frac{d q^i}{ds} \frac{d q^j}{ds},
\end{align}
where $g^{ij}(\bm{q})$ is the inverse matrix element of the metric $g(\bm{q})$.
Then Fermat's principle for anisotropic media is given by  
\begin{align}
  0 = \delta \int_A^B n(\bm{q}(s)) \, \sqrt{ \frac{d \bm{q}}{ds} \cdot  \frac{d \bm{q}}{ds} } \, ds,
     \label{FermatP}
\end{align}
where the dot product is defined with respect to $g_{ij}(\bm{q})$, e.g., $\bm{q} \cdot \bm{v} := g_{ij}(\bm{q}) q^i v^j$.
The line element $ds$ of the arc-length parameter $s$ satisfies
\begin{align}
  (ds)^2 = g_{ij}(\bm{q}) dq^i dq^j,   \quad \Leftrightarrow \quad \left\vert \frac{d \bm{q}}{ds}  \right\vert^2 
  = \frac{d \bm{q}}{ds} \cdot  \frac{d \bm{q}}{ds}  =
  g_{ij}(\bm{q}) \frac{dq^i}{ds} \frac{d q^j}{ds}  =1.
  \label{dsdq}
 \end{align}
The corresponding Lagrangian is
\begin{align}
  L \left( \bm{q}, \frac{d \bm{q}}{ds} \right) = n(\bm{q}) \, \sqrt{ g_{ij}(\bm{q}) \frac{dq^i}{ds} \frac{d q^j}{ds} },
     \label{L}
\end{align}
from which the canonical momentum $\bm{p}(s)$ is given by
\begin{align}
   p_i := \frac{\partial L}{\partial \left(\frac{dq^i}{ds} \right) }  = \frac{n(\bm{q}) \, g_{ij}(\bm{q}) \frac{dq^j}{ds}}{\sqrt{ g_{ij}(\bm{q}) \frac{dq^i}{ds} \frac{d q^j}{ds} }}
   =n(\bm{q}) \, g_{ij}(\bm{q}) \frac{dq^j}{ds},
   \label{p}
   \end{align}
where \eqref{dsdq} is used.
By solving this relation for $d q^i / ds$, we have
\begin{align}
 \frac{d q^i}{ds}  =  \frac{g^{ij}(\bm{q}) \, p_j}{n(\bm{q}) }.
 \label{v}
\end{align}
Note that
\begin{align}
   p_i \frac{dq^i}{ds}  = n(\bm{q}) \, g_{ij}(\bm{q}) \frac{dq^i}{ds} \frac{dq^j}{ds} = n(\bm{q}),
\end{align}
and
\begin{align}
   \sqrt{g^{ij}(\bm{q}) p_i p_j}  = n(\bm{q}) \, \sqrt{g_{ij}(\bm{q}) \frac{dq^i}{ds} \frac{dq^j}{ds}} = n(\bm{q}).
   \label{n}
\end{align}

The associated Euler-Lagrange equation becomes
\begin{align}
   \frac{d p_i}{ds} = \frac{\partial L}{\partial q^i} =  \frac{n(\bm{q})}{2} \, \frac{\partial g_{jk}(\bm{q})}{\partial q^i} \frac{dq^j}{ds} \frac{dq^k}{ds} + \frac{\partial n(\bm{q})}{\partial q^i}.
   \label{ELeq}
\end{align}
By using the relation
\begin{align}
   \frac{\partial g_{jk}(\bm{q})}{\partial q^i} g^{k \ell}(\bm{q}) =  -g_{jk}(\bm{q}) \frac{\partial g^{k \ell}(\bm{q})}{\partial q^i},
   \label{delg2delginv}
\end{align}
which is obtained from $g_{jk}(\bm{q}) g^{k \ell}(\bm{q}) = \delta_j^{\ell}$,
the first term of right hand side in \eqref{ELeq} becomes
\begin{align}
\frac{n(\bm{q})}{2} \, \frac{\partial g_{jk}(\bm{q})}{\partial q^i} \frac{dq^j}{ds} \frac{dq^k}{ds} 
= - \frac{1}{2 n(\bm{q})} \, \frac{\partial g^{jk}(\bm{q})}{\partial q^i} p_j p_k = - \frac{\partial}{\partial q^i} \, \sqrt{g^{jk} p_j p_k} .
\label{2ndterm}
\end{align}
Consequently,  from \eqref{v}, \eqref{n}, \eqref{ELeq} and \eqref{2ndterm}, the set of Euler-Lagrange equations is equivalent to the following set of equations
\begin{subequations}
\begin{empheq}[left=\empheqlbrace]{align}
 \frac{ d q^i}{ds} &= \frac{\partial}{\partial p_i} \Big( \sqrt{g^{jk}(\bm{q}) p_j p_k} - n(\bm{q}) \Big), 
 \\
 \frac{ d p_i}{ds} &= -\frac{\partial}{\partial q^i} \Big( \sqrt{g^{jk}(\bm{q}) p_j p_k} - n(\bm{q}) \Big), 
 \end{empheq}
\end{subequations}
which is the set of Hamilton's equations for the Hamiltonian \footnote{It should be noted that in order to clarify the time parameter ($s$ in this case), we denote any time-independent Hamiltonian as, for example, $H(\bm{q}(s), \bm{p}(s) )$ throughout the paper, instead of the usual notation $H(\bm{q}, \bm{p} )$. 
}
\begin{align}
H(\bm{q}(s), \bm{p}(s) ) = \sqrt{g^{jk}(\bm{q}) p_j p_k} - n(\bm{q}).
 \label{H}
\end{align}
Here the negative refractive index $- n(\bm{q})$ can be regarded as a potential function $U(\bm{q})$.
Note that, from \eqref{n} it follows that the total energy $E$ of this Hamiltonian $H(\bm{q}, \bm{p} )$ is zero.

Next in order to transform a Hamiltonian which has a potential function $U(\bm{q})$, to a geodesic Hamiltonian, which has no potential function, one resorts to
the JM transformation \cite{CGG17} (see \ref{append} for the details).
For Hamiltonian \eqref{H},
since $E=0$ and $U(\bm{q}) = - n(\bm{q})$, the JM transformed Hamiltonian $\tilde{H}$ becomes
\begin{align}
 \tilde{H} (\bm{q}(\tau), \bm{p}(\tau) ):= \frac{\sqrt{g^{jk}(\bm{q}) p_j p_k}}{E-U(\bm{q})} = \frac{1}{ n(\bm{q})} \sqrt{g^{jk}(\bm{q}) p_j p_k},
 \label{tildeH}
\end{align}
and
\begin{align}
  d \tau := (E - U(\bm{q}) ) ds =  n(\bm{q}) ds,
  \label{dtau}
  \end{align}
which relates the time parameter $s$ describes the Hamilton-flows by the Hamiltonian  \eqref{H} to
the time parameter $\tau$ describes those by the (JM transformed) geodesic Hamiltonian \eqref{tildeH}.
Hamilton's equations of motion are
\begin{subequations}
\label{setHeq}
\begin{empheq}[left=\empheqlbrace]{align}
\frac{d q^i}{d \tau} &= \frac{\partial \tilde{H}}{\partial p_i} = \frac{1}{n^2(\bm{q}) } \, g^{ij}(\bm{q}) p_j,  \\
\frac{d p_i}{d \tau} &= -\frac{\partial \tilde{H}}{\partial q^i} 
= \frac{1}{n(\bm{q}) } \Big( -\frac{1}{2n(\bm{q}) } \frac{\partial g^{jk}(\bm{q})}{\partial q^i} p_j p_k + \frac{\partial n(\bm{q})}{\partial q^i} \Big).
\label{Heq}
\end{empheq}
\end{subequations}
By using \eqref{dtau}, these equations are equivalent to  \eqref{v} and  \eqref{ELeq}.
In terms of the JM metric
\begin{align}
 \tilde{g}^{ij}(\bm{q}) :=  \frac{g^{ij}(\bm{q})}{(E - U(\bm{q}))^2}  = \frac{ g^{ij}(\bm{q})}{n^2(\bm{q})},
\end{align}
the transformed Hamiltonian \eqref{tildeH} is expressed as
\begin{align}
\tilde{H}(\bm{q}(\tau), \bm{p}(\tau)) =\sqrt{\tilde{g}^{jk}(\bm{q}) p_j p_k},
\end{align}
and then equations \eqref{setHeq} are rewritten as
\begin{subequations}
\label{tau-Heq}
\begin{empheq}[left=\empheqlbrace]{align}
\frac{d q^i}{d \tau} &=  \tilde{g}^{ij}(\bm{q}) p_j,  \\
\frac{d p_i}{d \tau} &=  -\frac{1}{2} \frac{\partial \tilde{g}^{jk}(\bm{q})}{\partial q^i} p_j p_k. 
\end{empheq}
\end{subequations}
Note that the transformed Hamiltonian $\tilde{H}(\bm{q}, \bm{p})$ is a homogeneous function of first order in the variable $p_i$. This property is related to the reparametrization invariance \cite{B98}, which we shall discuss in the following.
We focus on the reparametrization of a time parameter of a flow.
Note that Fermat's principle \eqref{FermatP} is invariant against the reparametrization of time parameter.
The action integral in \eqref{FermatP} is invariant by changing $s$ to any time parameter.
Hence it is important to specify the relation to the time parameter $t$ which is used in the gradient-flow equations \eqref{gradEq_eta}.
Let us introduce the time parameter $t$ whose line element $dt$ satisfies 
\begin{align}
  d\tau = n^2(\bm{q}) \, dt.
  \label{dt}
\end{align}
Using this reparametrization from $\tau$ to $t$, Hamilton's equation \eqref{tau-Heq} are further rewritten as the following set of the differential eqations with respect to the time parameter $t$:
\begin{subequations}
\label{t-Heq}
\begin{empheq}[left=\empheqlbrace]{align}
\frac{d q^i}{d t} &=  g^{ij}(\bm{q}) p_j, 
\label{tdotq}\\
\frac{d p_i}{d t} & =  -\frac{1}{2} \frac{\partial g^{jk}(\bm{q})}{\partial q^i} p_j p_k + \frac{1}{2} \frac{\partial  n^2(\bm{q})}{\partial q^i},
\label{tdotp}
\end{empheq}
\end{subequations}
which are  Hamilton's equations for the following natural Hamiltonian \footnote{A Hamiltonian consists of the kinetic energy and a potential function is called a \textit{natural Hamiltonian}.}
\begin{align}
  \mathsf{H}(\bm{q}(t), \bm{p}(t)) = \frac{1}{2} g^{ij}(\bm{q}) p_i p_j - \frac{1}{2}  n^2(\bm{q}).
\end{align}
By using \eqref{D}, \eqref{dtau} and \eqref{dt},
we finally obtain the anisotropic Huygens equation \eqref{anisoHuyEq} with respect to this time parameter $t$ as
\begin{align}
   \frac{\partial S(\bm{q} )}{\partial q^i} =  g_{ij}(\bm{q}) \, \frac{d q^j}{dt}, \quad i, j = 1,2, \ldots, m.
   \label{t-anisoHuyEq}
\end{align}

\section{Geodesic Hamiltonian in information geometry}
\label{sec:3}
Having found the relation  between Huygens' equation in terms of the time parameter $t$ and the transformed geodesic Hamiltonian in geometric optics in the previous section, we here consider a geodesic Hamiltonian in IG.

In previous works \cite{WSM21,WMS15}, the possible correspondence between IG and analytical mechanics are shown
and they are summarized as follows. 
\begin{subequations}
\label{correspondence}
\begin{empheq}[left=\empheqlbrace]{align}
 \textrm{$\eta$-coordinates: } \quad \eta_i  \quad &\Leftrightarrow \quad  q^i,  \\
 \textrm{$\theta$-coordinates: } \quad \theta^i  \quad &\Leftrightarrow \quad  -p_i.
\end{empheq}
\end{subequations}
Here $q^i$ and $p_i$ denote $i$-th component of the position and those of the momentum, respectively.
It is well known in IG that the $\eta$-potential function $ \Psi^{\star}(\bm{\eta})$ satisfies
\begin{align}
  d \Psi^{\star}(\bm{\eta}) = \theta^i \, d\eta_i,  \quad \textrm{or} \quad  \theta^i = \frac{\partial \Psi^{\star}(\bm{\eta})}{\partial \eta_i}.
  \label{dPsi_star}
\end{align}
Now introducing the arc-length parameter $s$, whose line element $ds$ satisfies
\begin{align}
   \big( ds \big)^2 = g^{jk}(\bm{\eta}) \, d\eta_j d\eta_k.
   \label{ds}
\end{align}
The gradient-flow equation \eqref{gradEq_eta} can be written in the following form
\begin{align}
   \frac{d \eta_i}{dt} = -g_{ij}(\bm{\eta}) \, \frac{\partial \Psi^{\star}(\bm{\eta})}{\partial \eta_j} = \nabla_{\eta} \Big(-\Psi^{\star}(\bm{\eta}) \Big),
    \label{gradEq-eta}
\end{align}
where the last term represents the natural gradient \cite{Amari98} of the negative $\eta$-potential, which is the entropy of a statistical model.
Our key idea is to regard this gradient relation \eqref{gradEq-eta} as an anisotropic Huygens equation \eqref{t-anisoHuyEq} in terms of the time-parameter $t$, and the level set of wave front $S(\bm{q})$ is regarded
as the level set of $- \Psi^{\star}(\bm{\eta})$.
Then the corresponding refractive index $n(\bm{\eta})$ is given by
\begin{align}
 n^2(\bm{\eta}) &= \big\vert  -\nabla_{\eta} \Psi^{\star}(\bm{\eta})  \big\vert^2 
 = g^{ij}(\bm{\eta}) \left(- g_{ik}(\bm{\eta}) \frac{\partial \Psi^{\star}(\bm{\eta})}{\partial \eta_k} \right) \left(- g_{j\ell}(\bm{\eta}) \frac{\partial \Psi^{\star}(\bm{\eta})}{\partial \eta_{\ell}} \right)  \notag \\
 & =g_{ij} (\bm{\eta}) \, \theta^i(\bm{\eta}) \theta^j(\bm{\eta}),
 \label{n2}
\end{align}
where \eqref{dPsi_star} is used. From \eqref{gradEq-eta} this relation is also expressed as
\begin{align}
 n^2(\bm{\eta}) = g^{ij}(\bm{\eta}) \frac{d \eta_i}{dt} \frac{d \eta_j}{dt} =  g^{ij}(\bm{\eta}) \frac{d \eta_i}{ds} \frac{d \eta_j}{ds} \left(\frac{ds}{dt} \right)^2
 = \left(\frac{ds}{dt} \right)^2,
 \label{n2-dsdt}
\end{align}
where in the last step we used \eqref{ds}.
By using \eqref{gradEq-eta}, \eqref{n2}, and \eqref{n2-dsdt}, it follows that
\begin{align}
  \frac{ d \Psi^{\star}(\bm{\eta})}{dt} &= \frac{\partial \Psi^{\star}(\bm{\eta}) }{\partial \eta_i}  \frac{d \eta_i}{dt} = 
  \frac{\partial \Psi^{\star}(\bm{\eta}) }{\partial \eta_i} \left( - g_{ij}(\bm{\eta})\, \frac{\partial \Psi^{\star}(\bm{\eta}) }{\partial \eta_j} \right) \nonumber \\
  &= -g_{ij}(\bm{\eta}) \, \theta^i(\bm{\eta}) \theta^j(\bm{\eta}) = -n^2(\bm{\eta}).
  \label{dPhis2dt}
\end{align}
The explicit expression of $n(\bm{\eta})$ is determined after the statistical model is specified. It will be shown
in the next subsections for Gaussian and Gamma models.

In relation to \eqref{tildeH}, we propose the following geodesic Hamiltonian in IG
\begin{align}
   \tilde{H}^{\rm IG}(\bm{\eta}(\tau), \bm{\theta}(\tau)) :=  \sqrt{ \tilde{g}_{jk}(\bm{\eta}) \, \theta^j \theta^k},
  \label{tildeH-IG}
\end{align}
where
\begin{align}
 \tilde{g}_{jk}(\bm{\eta}) :=  \frac{ g_{jk}(\bm{\eta}) }{n^2(\bm{\eta})}, \quad d\tau = n(\bm{\eta}) \, ds = n^2(\bm{\eta}) \, dt,
 \label{tilde_g}
 \end{align}
 and the line element $d \tau$ satisfies
 \begin{align}
  ( d \tau)^2 =  \tilde{g}^{i j}(\bm \eta) d \eta_i d \eta_j.
  \label{dtau2}
 \end{align}
Hamilton's equations of motion for this geodesic Hamiltonian $\tilde{H}^{\rm IG}$ are
\begin{subequations}
\label{Hflow_IG}
\begin{empheq}[left=\empheqlbrace]{align}
   \frac{d \eta_i}{d \tau} &= -\frac{\partial \tilde{H}^{\rm IG}}{\partial \theta^i}
   = -\tilde{g}_{i j}(\bm{\eta}) \; \theta^{j}, 
   \label{etadot} \\
  \frac{d \theta^i}{d \tau} &= \frac{\partial\tilde{H}^{\rm IG}}{\partial \eta_i}
   = \frac{1}{2} \, \frac{\partial \tilde{g}_{j k}(\bm{\eta})}{\partial \eta_i} \, \theta^j \theta^k. 
   \label{thetadot}
\end{empheq}
\end{subequations}
Hamilton's equations \eqref{Hflow_IG} are rewritten by
\begin{subequations}
\label{Heqs_in_t}
\begin{empheq}[left=\empheqlbrace]{align}
   \frac{d \eta_i}{d t} &= - \frac{d \tau}{d t} \, \frac{\partial \tilde{H}^{\rm IG}}{\partial \theta^i}
   = -g_{i j}(\bm{\eta}) \; \theta^{j},
   \label{d_eta2dt} \\
  \frac{d \theta^i}{d t} &= \frac{d \tau}{d t} \, \frac{\partial\tilde{H}^{\rm IG}}{\partial \eta_i}
   = \frac{n^2(\bm{\eta})}{2 } \, \frac{\partial \tilde{g}_{j k}(\bm{\eta})}{\partial \eta_i} \, \theta^j \theta^k 
   \notag \\
   &= \frac{1}{2} \, \frac{\partial g_{j k}(\bm{\eta})}{\partial \eta_i} \, \theta^j \theta^k  -\frac{1}{2}  \frac{\partial n^2(\bm{\eta})}{\partial \eta_i},
\end{empheq}
\end{subequations}
in terms of the time-parameter $t$.
In this way, by regarding the gradient-flow equations \eqref{gradEq-eta} as anisotropic Huygens' equation, its dynamics is related to Hamilton's flow $(\eta_i(\tau), \theta^i(\tau))$ described by Hamilton's equations \eqref{Hflow_IG} of the proposed geodesic Hamiltonian \eqref{tildeH-IG}, and to Hamilton's flow $(\eta_i(t), \theta^i(t))$ described by Hamilton's equations \eqref{Heqs_in_t} of the natural Hamiltonian
\begin{align}
  H^{\rm IG}(q^i(t), p_i(t)) = \frac{1}{2} \, g_{j k}(\bm{\eta}) \, \theta^j \theta^k  -\frac{1}{2}  n^2(\bm{\eta}).
\end{align}


Next we shall relate Hamilton's equations \eqref{Hflow_IG} to the linear differential equations
(the first sets in \eqref{gradEq}).
From \eqref{etadot}, we have
\begin{align}
  \theta^i = - \tilde{g}^{ij}(\bm{\eta}) \, \frac{d \eta_j}{d \tau}.
  \end{align}
By utilizing this relation, we see that
\begin{align}
 \tilde{g}_{i j }(\bm{\eta}) \, \theta^i \theta^j &= \tilde{g}_{i j }(\bm{\eta}) \, \left(- \tilde{g}^{i k}(\bm{\eta}) \frac{d \eta_k}{d \tau} \right) \left(- \tilde{g}^{j \ell}(\bm{\eta}) \frac{d \eta_{\ell}}{d \tau} \right) \notag \\
 &= \tilde{g}^{i j}(\bm{\eta}) \frac{d \eta_i}{d \tau} \frac{d \eta_j}{d \tau} = \frac{\tilde{g}^{i j}(\bm{\eta}) d \eta_i d \eta_j}{(d \tau)^2}= 1,
\end{align} 
where in the last step we used \eqref{dtau2}.
By differentiating both sides of this relation $  \tilde{g}_{j k}(\bm{\eta}) \, \theta^j \theta^k =1$ with respect to $\eta_i$ we have
 \begin{align}
  \frac{\partial \tilde{g}_{j k }(\bm{\eta}) }{\partial \eta_i} \, \theta^j \theta^k + 2 \tilde{g}_{ j k }(\bm{\eta}) \, \frac{\partial \theta^j}{\partial \eta_i} \left(- \tilde{g}^{k \ell}(\bm{\eta}) \frac{d \eta_{\ell}}{d \tau} \right) =  \frac{\partial \tilde{g}_{j k }(\bm{\eta}) }{\partial \eta_i} \, \theta^j \theta^k - 2 g^{i j}(\bm{\eta})  \frac{d \eta_j}{d \tau}  = 0,
\end{align}
where we used $g^{i j}(\bm{\eta}) = \partial \theta^j / \partial \eta_i$. Consequently, we obtain
\begin{align}
 \frac{1}{2 } \, \frac{\partial \tilde{g}_{j k}(\bm{\eta})}{\partial \eta_i} \,  \theta^j \theta^k = g^{i j}(\bm{\eta})  \frac{d \eta_j}{d \tau}.
\end{align}
By using this relation in the right hand side of  \eqref{thetadot}, we confirm that
\begin{align}
 \frac{d \theta^i}{d \tau} = \frac{1}{2 } \, \frac{\partial \tilde{g}_{j k}(\bm{\eta})}{\partial \eta_i} \,  \theta^j \theta^k = g^{i j}(\bm{\eta})  \frac{d \eta_j}{d \tau}.
 \end{align}
By using  \eqref{etadot}, we finally obtain the first set of differential equations in \eqref{gradEq} as follows:
\begin{align}
  \frac{d \theta^i}{d t} &= \frac{d \tau}{d t} \, \frac{d \theta^i }{d \tau}  =  n^2(\bm{\eta}) \, g^{i j}(\bm{\eta}) \;  \frac{d \eta_j}{d \tau}
  = \tilde{g}^{i j}(\bm{\eta}) \;  \frac{d \eta_j}{d \tau}
  = - \tilde{g}^{i j}(\bm{\eta}) \;  \tilde{g}_{j k}(\bm{\eta}) \, \theta^k \notag \\
  &= - \delta^i_k \, \theta^k = -\theta^i.
\end{align}

\subsection{Gaussian model}
\label{Gaussian}
As a simple example, we here consider the Gaussian model, or Normal $N(\mu, \sigma)$ model, which is well known \cite{Amari} in IG and the pdf is given by
\begin{align}
 p_{\rm G}(x; \mu, \sigma) = \frac{1}{\sqrt{2 \pi \, \sigma^2}} \,
     \exp \left[ -\frac{(x-\mu)^2}{2 \sigma^2} \right],
\end{align}
where $\mu$ denotes the mean and $\sigma^2$ is the dispersion. 
It is also known that the associated natural $\theta$-coordinates and $\eta$-coordinates \cite{Amari} are
\begin{align}
 \theta^1 = \frac{\mu}{\sigma^2}, \quad \theta^2 = -\frac{1}{2 \sigma^2}, \quad
 \eta_1 = \mu, \quad \eta_2 = \mu^2 + \sigma^2,
\end{align}
where one should not confuse the superscript in $\theta$ variables  with exponents.
The $\eta$-potential $\Psi^{\star}_{\rm G}(\bm{\eta})$ is
\begin{align}
 \Psi^{\star}_{\rm G}(\bm{\eta}) = - \frac{1}{2} \ln (2 \pi \mathrm{e}) - \frac{1}{2}\ln \left(\eta_2 - (\eta_1)^2  \right)
  =- \frac{1}{2} \ln (2 \pi \mathrm{e})  - \frac{1}{2}\ln \sigma^2.
\end{align}
The components $g_{ij}(\bm{\eta})$ of the metric tensor $g(\bm{\eta})$ are 
\begin{align}
g_{ij}(\bm{\eta}) =
 2 \big( \eta_2 - (\eta_1)^2 \big)
\left( \begin{array}{cc}
   \frac{1}{2} & \eta_1 \\[1ex] 
    \eta_1  &  ( \eta_1)^2 + \eta_2
\end{array}
\right)
=
2 \sigma^2
\left( \begin{array}{cc}
   \frac{1}{2} & \mu  \\[1ex] 
    \mu  &  2 \mu^2 + \sigma^2
\end{array}
\right)
.
\end{align}
After straightforward calculations, we obtain that
\begin{align}
   n_{\rm G} :=  \sqrt{ g_{j k}(\bm{\eta}) \theta^j \theta^k} 
   = \frac{1}{\sqrt{2}}.
\end{align}
Note that  $n_{\rm G}$ does not depend on $\bm{\eta}$, we hence denote this quantity as $n_{\rm G}$ instead of $n_{\rm G}(\bm{\eta})$.

From \eqref{d_eta2dt}, we have
\begin{align}
\frac{d \eta_1}{d t} = \frac{d \mu}{d t} = -g_{11}(\bm{\eta}) \, \theta^1 - g_{12}(\bm{\eta}) \, \theta^2  = 0, 
\label{dmu2dt}
\end{align}
and
\begin{align}
 \frac{d \eta_2}{d t} &= \frac{d}{d t} \big( \mu^2 + \sigma^2 \big) 
  = -g_{21}(\bm{\eta}) \, \theta^1 - g_{22}(\bm{\eta}) \, \theta^2 \nonumber \\
  &= -2 \sigma^2 \mu \frac{\mu}{\sigma^2} + 2 \sigma^2 (2\mu^2+ \sigma^2)\frac{1}{2 \sigma^2}  
  =  \sigma^2 = \eta_2 - (\eta_1)^2.
\end{align}
Next, the $t$-derivative of the $\eta$-potential becomes
\begin{align}
   \frac{d \Psi^{\star}_{\rm G}(\bm{\eta})}{dt} &= \theta^i  \frac{d \eta_i}{dt} = \frac{\mu}{\sigma^2} \frac{d \mu}{dt} - \frac{1}{2 \sigma^2} \frac{d( \mu^2 + \sigma^2)}{dt}  
    = -\frac{1}{2 \sigma^2} \frac{d \sigma^2}{dt} = - (n_{\rm G})^2,
\end{align}
where in the last step \eqref{dPhis2dt} is used. Consequently, we obtain the explicit expression of $dt$ as 
\begin{align}
    d t = \frac{d \sigma^2}{2 \sigma^2 (n_{\rm G})^2 } = \frac{d \sigma^2}{\sigma^2}.
\end{align}
We then confirm the gradient-flow equations \eqref{gradEq} are indeed satisfied as follows.
\begin{subequations}
\begin{empheq}[left=\empheqlbrace]{align}
   \frac{d}{d t} \theta^1 &= \sigma^2 \frac{d}{d \sigma^2} \left( \frac{\mu}{\sigma^2} \right) = -\frac{\mu}{\sigma^2} = - \theta^1, \\
\frac{d}{d t} \theta^2 &= \sigma^2 \frac{d}{d \sigma^2} \left(- \frac{1}{2 \sigma^2} \right) = \frac{1}{2 \sigma^2} = - \theta^2. 
\end{empheq}
\end{subequations}
Note that from \eqref{dmu2dt}, we see that
the parameter $\mu$ is independent of $t$. Consequently the line element $ds$ satisfies
\begin{align}
 \big(ds \big)^2 =  \left( n_{\rm G} \, dt \right)^2 = \frac{ (d\mu)^2 + 2 (d \sigma)^2}{\sigma^2} = \frac{ 2 (d \sigma)^2}{\sigma^2},
\end{align}
from which we confirm
\begin{align}
  d s  = \sqrt{2} \,  \frac{d \sigma}{\sigma} = \frac{1}{\sqrt{2}} \frac{d \sigma^2}{\sigma^2} = n_{\rm G} \,  d \ln \sigma^2 = n_{\rm G} \, dt.
  \label{ds_G}
\end{align}
Recall that the geodesics of a null metric in a flat space are the trajectories of light rays.
From Eq. \eqref{ds_G},  we obtain the null (light-like) metric \cite{CW22} in a flat space as follows.
\begin{align}
  0 = - (n_{\rm G})^2 \, (dt)^2 + \frac{ 2 (d \sigma)^2}{\sigma^2} 
  =  \frac{1}{\sigma^2}  \left\{ - (n_{\rm G})^2 \sigma^2 \, (dt)^2 +  2 (d \sigma)^2   \right\}.
\end{align}

\subsection{Gamma model}
\label{Gamma}
As another example, the Gamma model is considered here.
The pdf is given by
\begin{align}
 p_{\rm \Gamma}(x; \beta, \nu) &= x^{\nu-1} \frac{\beta^{\nu}}{\Gamma(\nu)} \,
     \exp\big[ -\beta x \big] \nonumber \\
     &= \exp \Big[-\beta x + (\nu-1) \ln x - \ln \Gamma(\nu) + \nu \ln \beta \Big],
\end{align}
where $\Gamma(\nu)$ is the gamma function, $\beta$ is the inverse scale parameter and $\nu$ is the shape parameter. 
The associated natural $\theta$-coordinates and $\eta$-coordinates are
\begin{align}
 \theta^1 = -\beta, \quad \theta^2 =  \nu - 1, \quad
 \eta_1 =  \frac{\nu}{\beta}, \quad \eta_2 = \phi(\nu)-\ln \beta,
\end{align}
where $\phi(\nu) := (d / d\nu) \ln \Gamma(\nu)$ is the digamma function.
The components $g_{ij}(\bm{\eta})$ and $g^{ij}(\bm{\eta})$ of the metric tensor $g(\bm{\eta})$ are 
\begin{align}
g_{ij}(\bm{\eta}) =
\left( \begin{array}{cc}
   \frac{\nu}{\beta^2} & \frac{1}{\beta} \\[1ex] 
   \frac{1}{\beta}   & \phi'(\nu)
\end{array}
\right)
, \quad
g^{ij}(\bm{\eta}) =
\frac{1}{\nu \phi'(\nu) - 1}
\left( \begin{array}{cc}
   \beta^2 \phi'(\nu) & -\beta \\[1ex] 
   -\beta   & \nu
\end{array}
\right),
\end{align}
where $\phi'(\nu) := d \phi(\nu) / d \nu$.
After straightforward calculations, we obtain that
\begin{align}
   n_{\rm \Gamma}(\nu) := \sqrt{g_{ij}(\bm{\eta}) \, \theta^i \theta^j } = \sqrt{2-\nu + \phi'(\nu)(\nu-1)^2}.
\end{align} 
In this case the refractive index depends on the parameter $\nu$ only, hence we denote it as $n_{\rm \Gamma}(\nu)$.

Now, from \eqref{etadot}, we have
\begin{subequations}
\begin{empheq}[left=\empheqlbrace]{align}
  \frac{d \eta_1}{d t} &= \frac{d}{d t} \left( \frac{\nu}{\beta} \right) 
  = - g_{11}\, \theta^1 - g_{12} \, \theta^2 = 
  \frac{\nu}{\beta} - \frac{(\nu-1)}{\beta}  = \frac{1}{ \beta}, \\
  \frac{d \eta_2}{d t} &= \frac{d}{d t} \left( \phi(\nu)-\ln \beta \right) =
  - g_{21}\, \theta^1 - g_{22} \, \theta^2  
  = 1 - \phi'(\nu) (\nu-1), 
  \end{empheq}
\end{subequations}
from which we obtain
\begin{align}
   \frac{1}{\beta} \frac{d \beta}{d t} = \frac{1}{\nu-1} \frac{d \nu}{d t}.
   \label{dbeta2dtau}
\end{align}
This means that $\beta$ and $\nu$ are not independent of each other. Then we choose $\beta$ as the independent variable and $\nu=\nu(\beta)$ as the dependent variable. Consequently, \eqref{dbeta2dtau} leads to
\begin{align}
   \frac{d \nu}{d \beta} = \frac{\nu-1}{\beta}.
   \label{dnu}
\end{align}
By taking $t$-derivative of the $\eta$-potential $\Psi^{\star}(\bm{\eta})$, we have
\begin{align}
   \frac{d \Psi^{\star}(\bm{\eta})}{dt} = \theta^i  \frac{d \eta_i}{dt} 
   = \Big\{ \phi'(\nu)(\nu-1)-1 \Big\} \frac{d \nu}{dt} +\frac{1}{\beta} \frac{d \beta}{dt}.
   \label{exp-dt}
\end{align}
By using \eqref{dnu},  Eq. \eqref{exp-dt} becomes
\begin{align}
  \frac{d \Psi^{\star}(\bm{\eta})}{dt} = n_{\rm \Gamma}^2(\nu) \, \frac{1}{\beta} \frac{d \beta}{dt}.
\end{align}
Comparing this to $d \Psi^{\star}(\bm{\eta}) / dt = - n_{\rm \Gamma}^2(s)$, we finally obtain that
\begin{align}
   dt =  -\frac{d \beta}{\beta} = -\frac{d \nu }{\nu-1},
\end{align}
where Eq. \eqref{dnu} is used.
We then confirm the gradient-flow equations \eqref{gradEq} 
are indeed satisfied as follows:
\begin{subequations}
\begin{empheq}[left=\empheqlbrace]{align}
   \frac{d}{dt} \theta^1 &= -\beta \frac{d}{d \beta} \left( -\beta \right) = \beta = - \theta^1, \\
\frac{d}{dt} \theta^2 &= -\beta \frac{d}{d \beta} \left(\nu-1 \right) =  -\beta \frac{d \nu}{d \beta} = \nu-1 = - \theta^2. 
\end{empheq}
\end{subequations}

The line-element $d s$ of the arc-length is computed as follows:
\begin{align}
 (d s)^2 = g^{ij} d\eta_i d\eta_j = \nu \left(\frac{d \beta}{\beta} \right)^2 - 2 d\nu \frac{d \beta}{\beta} + \phi'(\nu) (d\nu)^2,
\end{align}
which is rewritten in the form:
\begin{align}
 (d s)^2 = \nu \left\{ \left(\frac{d \beta}{\beta} \right)^2 - \left(\frac{d \nu}{\nu-1} \right)^2 \right\} + \nu \left(\frac{d \nu}{\nu-1} \right)^2 - 2 d\nu \frac{d \beta}{\beta} + \phi'(\nu) (d\nu)^2.
\end{align}
From the relation \eqref{dnu}, we see that the term in the curly brackets (or braces) is zero and we thus confirm
\begin{align}
 d s = \sqrt{ 2-\nu + \phi'(\nu) (\nu-1)^2  } \,  \frac{d \beta}{\beta}  = n_{\rm \Gamma}(\nu) \, \frac{d \beta}{\beta}.
\end{align}

\subsection{Relations to Replicator equations}
\label{RepEq}
Replicator equation plays a fundamental role in  mathematical  biology for describing evolutionary game dynamics and population dynamics.
The relation between replicator equations and IG was first pointed out by Harper \cite{Harper}. 
It is shown \cite{AE05} that replicator equations follow naturally from the exponential affine structure of the simplex in IG. 

Let us consider a replicator equation in the following form:
\begin{align}
  \frac{d}{dt} p_{\theta}(x, t) = -  p_{\theta}(x, t)  \Big( \ln  p_{\theta}(x, t) - {\rm E}_{ p_{\theta}} \left[ \ln  p_{\theta}(x, t) \right] \Big),
\label{RE}
\end{align}
where $ {\rm E}_{p_{\theta}} [ \cdot ]$ denotes the expectation value with respect to a pdf $p_{\theta}(x, t)$.
Suppose that the $\theta$-variables depend on the time parameter $t$ as $\theta^k = \theta^k(t)$ and that $ p_{\theta}(x, t) $ is an exponential pdf, i.e.,
\begin{align}
  p_{\theta}(x, t) = \exp \left[ \theta^k(t) F_k(x) - \Psi(\bm{\theta}(t)) \right].
\end{align}
Taking derivative of $p_{\theta}(x, t)$ with respect to $t$, we have
\begin{align}
 \frac{d}{dt} p_{\theta}(x, t) &= \left( \frac{d \theta^k(t)}{dt} \, F_k(x)  - \frac{\partial \Psi(\bm{\theta}(t))}{\partial \theta^k} \, \frac{d \theta^k(t)}{dt} \right) p_{\theta}(x, t) \nonumber \\ 
  &= \frac{d \theta^k}{dt} \, \Big( F_k(x)  - \eta_k(t)  \Big) p_{\theta}(x, t),
 \label{rel1}
\end{align}
where we used the relation $\partial \Psi(\bm{\theta}(t)) / \partial \theta^k = \eta_k (t)$.
Since
\begin{align}
  \ln p_{\theta}(x, t) - {\rm E}_{ p_{\theta}} \left[ \ln  p_{\theta}(x, t) \right] = \theta^k(t) \Big(F_k(x) -\eta_k(t) \Big),
\end{align}
the replicator equation \eqref{RE} for an exponential pdf $p_{\theta}(x, t)$ becomes
\begin{align}
  \frac{d}{dt} p_{\theta}(x, t) = -   \theta^k(t) \Big(F_k(x) -\eta_k(t) \Big) p_{\theta}(x, t).
  \label{rel2}
\end{align}
Hence, if \eqref{rel1} and \eqref{rel2} are equivalent to each other, we have
\begin{align}
  \frac{d \theta^k(t)}{dt} = -\theta^k(t),
  \label{gradEq1}
\end{align}
which is the first set of the linear differential equations in \eqref{gradEq}.
Conversely, if \eqref{gradEq1} is satisfied, the derivative \eqref{rel1} of an exponential pdf $p_{\theta}(x, t)$ is equivalent to
the replicator equation \eqref{RE}.

\subsection{The gradient-flow equations associated with the $\theta$-potential function}
\label{2ndset}

Here, we'd like to make a short discussion on the second set of the gradient-flow equations in \eqref{gradEq_eta}.
They are associated with the $\theta$-potential function $\Psi(\bm{\theta})$, which is
the total Legendre transformation of the $\eta$-potential function $\Psi^{\star}(\bm{\eta})$.
It is well known in IG \cite{Amari} that the $\theta$-potential function $ \Psi(\bm{\theta})$ satisfies
\begin{align}
  d \Psi(\bm{\theta}) = \eta_i \, d\theta^i,  \quad \textrm{or} \quad  \eta_i = \frac{\partial \Psi(\bm{\theta})}{\partial \theta^i}.
 \label{dPsi}
\end{align}
The line element $ds$ of the arc-length parameter $s$ satisfies
\begin{align}
   \big( ds \big)^2 = g_{jk}(\bm{\theta}) \, d\theta^j d\theta^k,
   \label{ds2}
\end{align}
which is of course same as the line element $ds$ in \eqref{ds}.
The second set of gradient-flow equations \eqref{gradEq_eta} can be written in the following form
\begin{align}
   \frac{d \theta^i}{dt} = g^{ij}(\bm{\theta}) \, \frac{\partial \Psi(\bm{\theta})}{\partial \theta^j} = \nabla_{\theta} \Psi(\bm{\theta}),
    \label{gradEq-theta}
\end{align}
where the last term represents the natural gradient \cite{Amari98} of the $\theta$-potential function  $\Psi(\bm{\theta})$.
We regard this gradient relation \eqref{gradEq-theta} as an anisotropic Huygens equation \eqref{t-anisoHuyEq} in terms of the evolution parameter $t$, in which the level set of wave front $S^{\star}(\bm{q})$ is regarded
as the level set of $\Psi(\bm{\theta})$.
Then, the corresponding refractive index $n^{\star}(\bm{\theta})$ is given by
\begin{align}
 \left( n^{\star}(\bm{\theta}) \right)^2 = \big\vert  \nabla_{\theta} \Psi(\bm{\theta})  \big\vert^2 
 = g_{ij}(\bm{\theta}) g^{ik}(\bm{\theta}) \frac{\partial \Psi}{\partial \theta^j} \, g^{k\ell}(\bm{\theta}) \frac{\partial \Psi}{\partial \theta^{\ell}}    =g^{ij}(\bm{\theta})(\bm{\theta}) \eta_i(\bm{\theta}) \eta_j(\bm{\theta}),
 \label{n*2}
\end{align}
where Eq. \eqref{dPsi} is used. This relation is also expressed as
\begin{align}
  \left( n^{\star}(\bm{\theta}) \right)^2 = g_{ij}(\bm{\theta}) \frac{d \theta^i}{dt} \frac{d \theta^j}{dt} =  g_{ij}(\bm{\theta}) \frac{d \theta^i}{ds} \frac{d \theta^j}{ds} \left(\frac{ds}{dt} \right)^2
 = \left(\frac{ds}{dt} \right)^2,
 \label{n*2-dsdt}
\end{align}
where in the last step we used Eq.~\eqref{ds2}.
By using \eqref{gradEq-theta}, \eqref{n*2}, and \eqref{n*2-dsdt}, it follows that
\begin{align}
  \frac{ d \Psi(\bm{\theta})}{dt} &= \frac{\partial \Psi(\bm{\theta}) }{\partial \theta^i}  \frac{d \theta^i}{dt} = 
  \frac{\partial \Psi(\bm{\theta}) }{\partial \theta^i}  g^{ij}(\bm{\theta}) \frac{\partial \Psi(\bm{\theta}) }{\partial \theta^j}  
  = g^{ij}(\bm{\theta}) \eta_i(\bm{\theta}) \eta_j(\bm{\theta}) =  \left( n^{\star}(\bm{\theta}) \right)^2.
\end{align}
As same as in the case of $n(\bm{\eta})$, the explicit expression of $n^{\star}(\bm{\theta})$ is determined after the statistical model is specified.

With these settings, we propose that the geodesic Hamiltonian in IG is
\begin{align}
   \tilde{H}^{\star}_{\rm IG}(\bm{\eta}(\tau^{\star}), \bm{\theta}(\tau^{\star})) :=  \sqrt{ \tilde{g}^{jk}(\bm{\theta}) \, \eta_j \eta_k},
  \label{tildeH*-IG}
\end{align}
where
\begin{align}
 \tilde{g}^{jk}(\bm{\theta}) :=  \frac{ g^{jk}(\bm{\theta}) }{\left( n^{\star}(\bm{\theta}) \right)^2}, \quad d\tau^{\star} = n^{\star}(\bm{\theta}) \, ds = \left( n^{\star}(\bm{\theta}) \right)^2 \, dt.
 \end{align}
Hamilton's equations for this transformed $\tilde{H}^{\star}_{\rm IG}$ are
\begin{subequations}
\begin{empheq}[left=\empheqlbrace]{align}
   \frac{d \eta_i}{d \tau^{\star}} &= -\frac{\partial \tilde{H}^{\star}_{\rm IG}}{\partial \theta^i}
   = -\frac{1}{2} \, \frac{\partial \tilde{g}^{j k}(\bm{\theta})}{\partial \theta^i} \, \eta_j \eta_k,
   \\
  \frac{d \theta^i}{d \tau^{\star}} &= \frac{\partial\tilde{H}^{\star}_{\rm IG}}{\partial \eta_i}
   =  \tilde{g}^{i j}(\bm{\theta}) \; \eta_{j}.
\end{empheq}
     \label{H*flow_IG}
\end{subequations}
which are equivalent to
\begin{subequations}
\begin{empheq}[left=\empheqlbrace]{align}
   \frac{d \eta_i}{d t} &= - \frac{d\tau^{\star}}{d t} \, \frac{\partial \tilde{H}^{\star}_{\rm IG}}{\partial \theta^i}
   = -\frac{1}{2} \, \frac{\partial g^{j k}(\bm{\theta})}{\partial \theta^i} \, \eta_j \eta_k  +\frac{1}{2}  \frac{\partial \left( n^{\star}(\bm{\theta}) \right)^2}{\partial \theta^i},
   \\
  \frac{d \theta^i}{d t} &= \frac{d \tau^{\star}}{d t} \, \frac{\partial\tilde{H}^{\star}_{\rm IG}}{\partial \eta_i}
       = \left( n^{\star}(\bm{\theta}) \right)^2 \, \tilde{g}^{i j}(\bm{\theta}) \, \eta_{j} = g^{i j}(\bm{\theta}) \, \eta_{j},
       \label{dthetai2dt}
       \end{empheq}
\end{subequations}
in terms of the time-parameter $t$.

As an example, we consider the Gaussian model discussed in \ref{Gaussian}.
The components $g^{ij}(\bm{\theta})$ of the metric tensor $g(\bm{\theta})$ are 
\begin{align}
g^{ij}(\bm{\theta}) = 2 \left( \begin{array}{cc}
   (\theta^1)^2 -\theta^2 & \theta^1 \theta^2 \\[1ex] 
   \theta^1 \theta^2   & (\theta^2)^2
   \end{array}
\right) =
\frac{1}{2 \sigma^4}
\left( \begin{array}{cc}
   4 \mu^2 + 2 \sigma^2 & -2 \mu \\[1ex] 
   -2 \mu  & 1
   \end{array}
\right).
\end{align}
After straightforward calculations, we obtain that
\begin{align}
   n^{\star}_{\rm G}(\bm{\theta}) = \sqrt{ g^{11}(\bm{\theta})  (\eta_1)^2 + 2 g^{12}(\bm{\theta})  \eta_1 \eta_2+
  g^{22}(\bm{\theta})  (\eta_2)^2} = \sqrt{\frac{1}{2} \left(1+\frac{\mu^4}{\sigma^4}\right)}.
\end{align} 
By using Eq. \eqref{dthetai2dt}, we have
\begin{subequations}
\begin{empheq}[left=\empheqlbrace]{align}
  \frac{d \theta^1}{d t} &= \frac{d }{d t} \left( \frac{\mu}{\sigma^2} \right) 
  = \frac{d \mu }{d t}  \frac{1}{\sigma^2} - \frac{2 \mu}{\sigma^3}    \frac{d \sigma}{d t} 
  = g^{11}(\bm{\theta}) \, \eta_1 + g^{12}(\bm{\theta})  \, \eta_2 = \frac{\mu^3}{\sigma^4}, \\
  \frac{d \theta^2}{d t} 
  &= -\frac{d}{d t} \frac{1}{2 \sigma^2}
  = \frac{1}{ \sigma^3}   \frac{d\sigma}{d t} 
  = g^{21}(\bm{\theta}) \, \eta_1 + g^{22}(\bm{\theta})  \, \eta_2 
  = \frac{1}{2 \sigma^2} - \frac{\mu^2}{2 \sigma^4} . 
  \end{empheq}
\end{subequations}
which are rearranged as follows.
\begin{subequations}
\begin{empheq}[left=\empheqlbrace]{align}
  \frac{d \mu }{ \mu d t}&   - \frac{2}{\sigma}  \frac{d \sigma}{d t}  = \frac{\mu^2}{\sigma^2}, \\
  \frac{1}{ \sigma}   \frac{d\sigma}{d t} &= \frac{1}{2} \left(1 - \frac{\mu^2}{ \sigma^2} \right) . 
  \end{empheq}
\end{subequations}
By combining these relations we have
\begin{align}
\frac{1}{2}\left(  1- \frac{\mu^2}{\sigma^2} \right) \left(\frac{d \mu }{d t}  - \frac{2 \mu}{\sigma}    \frac{d \sigma}{d t} \right)
= \frac{\mu^3}{\sigma^3}  \frac{d\sigma}{d t},
\end{align}
from which it follows 
\begin{align}
\frac{d\sigma}{\sigma}
= \frac{1}{2} \left(  1 - \frac{\mu^2}{ \sigma^2} \right) \frac{d \mu }{\mu}.
\label{dsigmadmu}
\end{align}
This shows that the relation between the flows $\sigma(t)$ and $\mu(t)$.

Next, by taking derivative of the $\theta$-potential $\Psi(\bm{\theta}) $ with respect to $t$, we have
\begin{align}
   \frac{d \Psi(\bm{\theta})}{dt} &=  \eta_i \, \frac{d \theta^i}{dt} 
   = \frac{\mu^2}{\sigma^2} \frac{d \mu}{\mu dt} + \left(1-\frac{\mu^2}{\sigma^2}\right) \frac{d \sigma}{\sigma dt}      
   =    \frac{1}{2}   \left(1+\frac{\mu^4}{\sigma^4}\right) \frac{d \mu}{\mu dt}.
\end{align}
Comparing this to  $d \Psi(\bm{\theta}) / dt = (n^{\star}_{\rm G} )^2$, it is found that
\begin{align}
   dt =  d \ln \mu.
\end{align}
We then confirm the second set of the gradient-flow equations in \eqref{gradEq} are indeed satisfied as follows:
\begin{subequations}
\begin{empheq}[left=\empheqlbrace]{align}
   \frac{d}{d t} \eta_1 &= \mu \, \frac{d}{d \mu} \mu = \mu = \eta_1, \\
\frac{d}{d t} \eta_2 &= \mu \, \frac{d}{d \mu} \left(\mu^2 + \sigma^2 \right) 
= 2 \mu^2 + \frac{1}{2} \left(  1 - \frac{\mu^2}{ \sigma^2} \right) \sigma \frac{d \sigma^2}{d\sigma} = \eta_2,
\end{empheq}
\end{subequations}
where we used Eq. \eqref{dsigmadmu}.

\section{Conclusions}
\label{sec:conclusions}

It is well known in IG \cite{Amari} that affine-coordinates $\{ \theta^i \}$ and $\{ \eta_i \}$ are important and essential. However the evolution parameter $t$ in a solution, e.g., $\eta_i(t)$, of the gradient-flow equation \eqref{gradEq_eta} is not an affine-parameter. In order to describe Hamilton-flow or gradient-flow, it is important to specify the evolution parameter $t$ because it is conjugate  to the energy $E$ which is the value of the Hamiltonian.  For the geodesic Hamiltonian  $\tilde{H}^{\rm IG}(\bm{\eta}(\tau), \bm{\theta}(\tau))$, the corresponding evolution parameter is $\tau$ which is the arc-length parameter associated with the JM transformed metric $\tilde{g}$. It is not trivial to relate these different evolution parameters in general. 
We have revisited the relation between Hamilton flows and gradient flows in IG, and specified the evolution parameter $t$ in the gradient-flow equation in IG by relating $t$ to the arc-length parameter $\tau$ associated to our proposed geodesic Hamiltonian  $\tilde{H}^{\rm IG}(\bm{\eta}(\tau), \bm{\theta}(\tau))$ in \eqref{tildeH-IG}.
We have shown how
such dynamical equations can be derived from the static (or equilibrium) relations in the frame work of IG.
By regarding a gradient-flow equation in IG as Huygens' equation in geometric optics,
we have related the gradient flow in IG to the geodesic flow induced by the geodesic Hamiltonian \eqref{tildeH-IG} in Riemannian geometry.
In this way, we have introduced a variational principle (Fermat's principle) in IG.
As concrete examples, we have shown the explicit expressions of $dt$ in terms of the parameters of the pdfs for Gaussian and Gamma models.
We hope that our results are useful for the further studies of dynamical problems in different fields, such as
thermodynamics \cite{WSM21}, statistical physics, cosmology, thermodynamics of black holes \cite{CW22}, and so on.
A further study of the gradient-flow equations in IG from the perspective of Wely geometry is preformed \cite{W23}.

As a by-product, the relation between the gradient-flow equations and  the replicator equations is pointed out in Section 3.  It is interesting for a future study whether the obtained relation for the exponential family can be extended for a generalized exponential family \cite{Harper11}.

\section*{Acknowledgments}
The first named author (T.W.) thanks to late Prof. M. Tanaka (Fukuoka University) for especially his interest to our works in the early stage.
The first named author (T.W.) was supported by Japan Society for the Promotion of Science (JSPS) Grants-in-Aid for Scientific Research (KAKENHI) Grant Number JP22K03431.
 The third named author (H. M.) was partially supported by the JSPS Grants-in-Aid for Scientific Research (KAKENHI) Grant Number JP19K03489 and JP23K03088.
\appendix

\section{Maupertuis' principle and Jacobi transformation}
\label{append}

Here we briefly review the Jacobi-Maupertuis (JM) transformation \cite{CGG17} and its associated metric.

In analytical mechanics, the dynamical path is determined from the variational principle of the action integral
$S_a$
\begin{align}
  0 = \delta S_a = \delta \int_{t_a}^{t_b} dt L \left( \bm{q},  \frac{d \bm{q}}{dt} \right)
  = \delta \int_{t_a}^{t_b} dt \left[ p_i \frac{dq^i}{dt} - H( \bm{q},  \bm{p} ) \right],
\end{align}
where $ L \left( \bm{q},  \frac{d \bm{q}}{dt} \right)$ and $H( \bm{q},  \bm{p} )$ are Lagrangian and Hamiltonian
of a system, respectively.
According to Maupertuis' principle, the dynamical path obtained from the variational principle of the action $S_a$
coincides with that obtained from the variational principle of the reduced action $S_0$
\begin{align}
  S_0 := \int_{t_a}^{t_b} dt \, p_i \frac{dq^i}{dt},
  \label{S0}
\end{align}
for a fixed total energy $H( \bm{q},  \bm{p} ) = E$.
This can be seen from the following relation:
\begin{align}
  0 = \delta S_a = \delta S_0 - \delta \int_{t_a}^{t_b} dt E = \delta S_0.
\end{align}

Now  consider the following relativistic Hamilton function $H(\bm{q}, \bm{p})$ given by
\begin{align}
  H(\bm{q}(t), \bm{p}(t)) = \sqrt{g^{jk}(\bm{q}) \, p_j p_k} + U(\bm{q}),
\end{align}
with $m$-degree of freedom. Here $g^{ij}(\bm{q})$ is the inverse matrix of a metric components $g_{ij}(\bm{q})$ of a smooth $m$-dimensinal Riemannian manifold, $U(\bm{q})$ denote a potential energy, and the total energy $E$ of this conservative system is constant.
From Hamilton's equations for this $ H(\bm{q}(t), \bm{p}(t))$, we have
\begin{align}
  \frac{d q^i}{dt} = \frac{\partial H}{\partial p_i} = \frac{g^{ij}(\bm{q}) \,  p_j}{ \sqrt{g^{k \ell}(\bm{q}) \, p_k p_{\ell}} }.
\end{align}
Then the reduced action \eqref{S0} is expressed as
\begin{align}
  S_0 = \int_{t_a}^{t_b} dt \, p_i \frac{\partial H}{\partial p_i} = \int_{t_a}^{t_b} dt \, \frac{g^{ij}(\bm{q})  \, p_i p_j}{\sqrt{g^{k \ell}(\bm{q}) \, p_k p_{\ell}} }
  = \int_{t_a}^{t_b} dt \, \sqrt{g^{jk}(\bm{q}) \, p_j p_k}.
\end{align}
On the fixed energy surface $ \mathcal{M}^{2m-1} = \Big(  H(\bm{q}(t), \bm{p}(t)) = E \Big)$,
the integral trajectories, say $q^i = \gamma^i(t)$, of  $H(\bm{q}, \bm{p})$  coincide with the integral trajectories $q^i = \tilde{\gamma}(\tau)$ of the new Hamitonian
\begin{align}
  \tilde{H}(\bm{q}(\tau), \bm{p}(\tau)) := \frac{\sqrt{g^{jk}(\bm{q}) p_j p_k}}{E - U(\bm{q})},
  \label{JMH}
\end{align}
with the transformed time-parameter $\tau$ whose line element $d \tau$ is given by
\begin{align}
  d \tau = \big( E - U(\bm{q}) \big) \, dt.
\end{align}
The resulting transformed $\tilde{H}(\bm{q}(\tau), \bm{p}(\tau))$ is called JM transformation of $H(\bm{q}, \bm{p})$
and describes geodesic motion with respect to the JM metric given by
\begin{align}
  \tilde{g}_{ij}(\bm{q}) :=  \big( E - U(\bm{q}) \big) \,  g_{ij}(\bm{q}),
\end{align}
so that its inverse matrix is
\begin{align}
  \tilde{g}^{ij}(\bm{q}) =  \frac{g^{ij}(\bm{q})}{E - U(\bm{q})}.
\end{align}
By using the JM metric, the transformed JM Hamiltonian \eqref{JMH} is expressed as
\begin{align}
  \tilde{H}(\bm{q}(\tau), \bm{p}(\tau)) = \sqrt{ \tilde{g}^{jk}(\bm{q}) p_j p_k},
  \label{transJMH}
\end{align}
which is a homogeneous function of first order in the variable $\bm{p}$.
The transformed parameter $\tau$ is the arc-length parameter whose line element $d \tau$ satisfies
\begin{align}
  d \tau^2 =  \tilde{g}^{ij}(\bm{q}) \, dq^i dq^j.
\end{align}
The corresponding set of Hamiton's equations is
\begin{subequations}
\label{JM-Heq}
\begin{empheq}[left=\empheqlbrace]{align}
\frac{d q^i}{d \tau} & =  \tilde{g}^{ij}(\bm{q}) \, p_j, 
\label{p-tau} \\
\frac{d p_i}{d \tau} &=  -\frac{1}{2} \frac{\partial \tilde{g}^{ij}(\bm{q})}{\partial q^i} \, p_j p_k.
\end{empheq}
\end{subequations}
By solving \eqref{p-tau} for $p_i$, we have
\begin{align}
 p_i =  \tilde{g}_{ij}(\bm{q}) \, \frac{d q^j}{d\tau}.
\end{align}
Substituting this relation into the transformed Hamiltonian \eqref{transJMH}, we see that
\begin{align}
  \sqrt{ \tilde{g}^{ij}(\bm{q}) \, p_i p_j}=\sqrt{ \tilde{g}_{jk}(\bm{q})  \, \frac{d q^i}{d\tau}  \frac{d q^j}{d\tau}} = 1.
\end{align}

\end{document}